\documentclass{kapproc} 
\let\footnote\savefootnote
\let\footnotetext\savefootnotetext
\let\footnoterule\savefootnoterule
\kluwerbib

\RequirePackage{graphicx}%
\RequirePackage{epsf}%

\begin{document}
\newcommand{\tvir}{T_{\rm vir}}
\newcommand{\vvir}{v_{\rm vir}}
\newcommand{\tmax}{T_{\rm max}}
\newcommand{\hmpc}{h^{-1}\,{\rm Mpc}}
\newcommand{\K}{\,{\rm K}}
\newcommand{\kms}{\,{\rm km}\,{\rm s}^{-1}}


\articletitle{How Do Galaxies Get Their Gas?}


\author{Neal Katz, Dusan Keres}

\affil{University of Massachusetts, Department of Astronomy,
Amherst, MA 01003}

\author{Romeel Dav\'e}

\affil{University of Arizona, Steward Observatory, Tucson, AZ 85721}

\author{David H. Weinberg}

\affil{Ohio State University,
Department of Astronomy, Columbus, OH 43210}


\begin{abstract}
Not the way one might have thought.  In hydrodynamic simulations of galaxy
formation, some gas follows the traditionally envisioned route, shock heating
to the halo virial temperature before cooling to the much lower temperature 
of the neutral ISM.  But most gas enters galaxies without ever heating close
to the virial temperature, gaining thermal energy from weak shocks and 
adiabatic compression, and radiating it just as quickly.  This ``cold mode'' 
accretion is channeled along filaments, while the conventional, ``hot mode'' 
accretion is quasi-spherical.  Cold mode accretion dominates high redshift
growth by a substantial factor, while at $z<1$ the overall accretion rate 
declines and hot mode accretion has greater relative importance.  The decline 
of the cosmic star formation rate at low $z$ is driven largely by geometry,
as the typical cross section of filaments begins to exceed that
of the galaxies at their intersections.
\end{abstract}

The conventional sketch of galaxy formation has its roots in classic
papers of the late '70s and early '80s, with the initial discussions
of collapse and cooling criteria by Binney (1977), Rees \& Ostriker (1977),
and Silk (1977), the addition of dark matter halos by White \& Rees (1978),
and the disk formation model of Fall \& Efstathiou (1980).
According to this sketch, gas falling into a dark matter potential
well is shock heated to approximately the halo virial temperature,
$\tvir = 10^6 (v_{\rm circ}/167\kms)^2 \K$.
Gas in the dense, inner regions of this shock heated halo radiates
its thermal energy, settles into a centrifugally supported disk,
and forms stars.
Mergers of disks can scatter stars onto disordered orbits,
producing spheroidal systems, which may regrow disks if they
experience subsequent gas accretion.
Over the last decade, the ideas of these seminal papers have been
updated and extended into a powerful ``semi-analytic'' framework
for galaxy formation calculations
(e.g., White \& Frenk 1991;
Kauffmann et al.\ 1993; Cole et al.\ 1994; Avila-Reese et al.\ 1998;
Mo, Mao, \& White 1998; Somerville \& Primack 1999).

The geometry seen in N-body and hydrodynamic cosmological simulations,
where the densest structures often form at the nodes of a filamentary
network, is clearly more complicated than the spherical geometry
envisioned in semi-analytic models.
Nonetheless, a substantial fraction of the gas in these simulations
does shock heat to
$T\sim \tvir$, and some of this gas does cool and settle into galaxies.
The approximate agreement between semi-analytic and smoothed particle
hydrodynamics (SPH) calculations of galaxy masses (e.g., Benson et al.
2001; Yoshida et al. 2002) has therefore been taken as evidence that
the conventional sketch, while idealized, captures most of the
essential physics.

There is, however, a long history of results suggesting that this outline
of the way that gas gets into galaxies is at best half of the story,
and perhaps the less important half.
Binney (1977) argued that the amount of shock heating could be
small for plausible physical conditions, with only a fraction of
the gas reaching temperatures $\sim \tvir$.
In the first SPH simulations of forming galaxies (Katz \& Gunn 1991),
which had idealized initial conditions but included small scale power
leading to hierarchical formation, most of the gas never heated above
$T\sim 3\times 10^4\K$, with much of the cooling radiation therefore
emerging in the Ly$\alpha$ line.
Katz \& White (1993) showed the importance of
filamentary structures as channels for gas accretion in simulations
with cold dark matter (CDM) initial conditions.

Two recent studies, based on SPH simulations of cosmological volumes,
reveal the situation even more starkly.
First, Fardal et al.\ (2001) show that most of the cooling radiation
in their simulations comes from gas with $T < 2\times 10^4\K$,
again implying that a significant fraction emerges in the Ly$\alpha$ line.
Since gas starting at $T\sim 10^6\K$ {\it must} radiate 90\% of its
thermal energy by the time it cools to $T\sim 10^5\K$, Fardal et al.\ (2001)
argue that the majority of the gas entering galaxies (indeed, the
majority of the gas experiencing any significant cooling) must not
be heated to the virial temperature of any dark matter halo resolved
by the simulation.
Second, Kay et al.\ (2000) directly track temperature histories
of particles that end up in their simulated galaxies, and they report that
only 11\% of these particles were ever heated above $10^5\K$.

We have begun to examine this issue more closely, using an SPH simulation
of a $22.222\hmpc$ box with an LCDM cosmology ($\Omega_m=0.4$,
$\Omega_\Lambda=0.6$, $n=0.95$, $h=0.65$, $\sigma_8=0.8$) and
$2\times 128^3$ particles.  The physical assumptions and numerical
techniques are described by Katz et al.\ (1996) and Dav\'e et al.\ (1997).
The resolution limit for galaxies, a baryon mass corresponding to
$64 m_{\rm SPH}$, is $6.7\times 10^9 M_\odot$ or a circular velocity of 
$v_{\rm vir}^{\rm min} = 46(1+z)^{1/2} \kms$. 
Figure~1 shows our key result.  
At four representative redshifts, we trace back each particle that
was recently added to a galaxy and record the maximum temperature
$\tmax$ that it had at any previous output.
Solid histograms show the distributions at each redshift in physical units,
demonstrating the decline of the overall accretion rate at low redshifts,
while the dotted histograms
are renormalized to better show the relative distribution.

\begin{figure*}
\epsscale{0.85}
\plotone{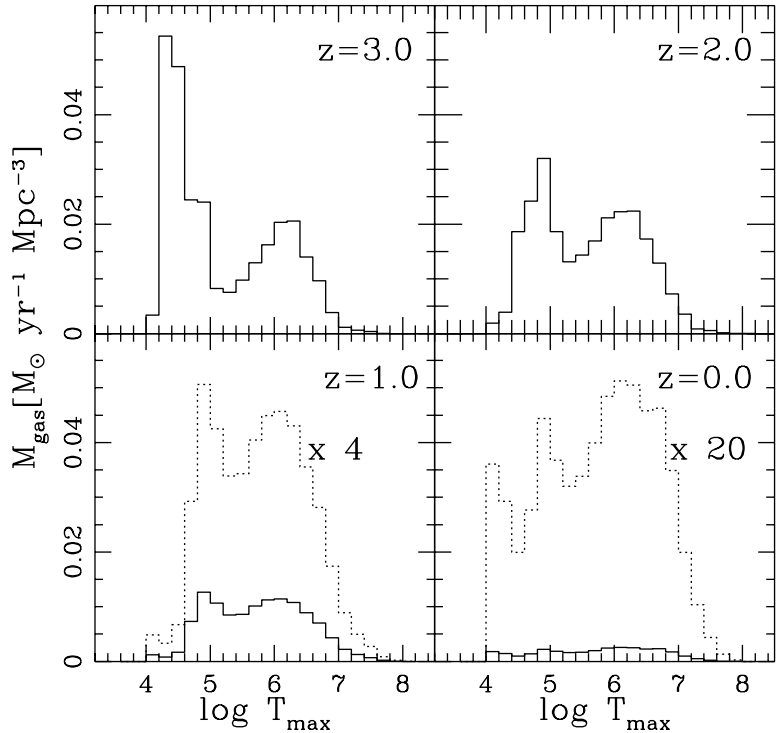}
\caption[]{The maximum temperature reached by gas accreted into galaxies
at the specified redshift.  The dotted line rescales the histogram to better
show the relative amount at different temperatures. Note the two distinct
modes of gas accretion, the ``cold mode'' and the ``hot mode''.
}
\end{figure*}

Our star formation and galaxy identification algorithms require
that gas particles cool to $T < 3\times 10^4\K$ and reach an overdensity of
$\rho/\bar{\rho} > 10^3$ before they are eligible to form stars
or be counted towards a galaxy's baryon mass component.
Figure~1 shows that gas reaches these physical conditions
via two distinct routes.
One of these corresponds to the conventional ``hot mode'' of gas accretion
outlined above, with gas heating to $T \geq \tvir \sim 10^5-10^7\K$ before
cooling.  In the second, ``cold mode'' of accretion, the maximum gas temperature
is $\tmax < \tvir$.
The cold mode dominates the total amount of accretion by a large factor
at high redshift, and
the hot mode becomes progressively more important at low redshift.
The characteristic $\tmax$ of the cold mode
rises with time, and the range of temperatures in each mode becomes broader,
so that they merge into a fairly continuous range of $\tmax$ values by $z=0$.
Overall, 42\% of the mass in galaxies at $z=0$ has 
$\tmax < T_{\rm vir}^{\rm min}$ when it was accreted, which is a
firm lower limit to the amount of ``cold mode'' accretion since most 
galaxy halos have $T_{\rm vir} > T_{\rm vir}^{\rm min}$.
Since our simulation outputs are spaced by $\Delta t \sim 0.3$ Gyr we
always underestimate a particle's true value of $\tmax$.
However, the cooling times are typically longer than $\Delta t$, and
the concordance of Figure~1 with the findings of Fardal et al. (2001),
whose cooling radiation argument applies to individual simulation outputs,
and of Kay et al.\ (2000), who examined every timestep in their simulation,
implies that this finite time resolution does not qualitatively change
our results.

What is going on?  One possibility, that the gas is cooling in halos
with low $\tvir$, can be immediately dismissed, since the lowest mass
halos resolved by the simulation have $\tvir \sim 76,000(1+z)\K$.
Furthermore, we have shown elsewhere (Murali et al.\ 2002)
that galaxies in these simulations gain most of their mass by smooth
accretion, not by mergers with pre-existing systems.
Even the more massive (and thus better resolved) galaxies in our
simulations are gaining most of their mass by accreting gas that
has never been hot. 

A second possibility is that this result is a numerical artifact:
gas that should be shock heated to high temperature is not.
We find qualitatively similar results at $z=3$ in a simulation whose
mass resolution is a factor of eight higher than the one illustrated
in Fig.~1, and at $z=0$ in a simulation whose mass resolution
is a factor of eight lower.  Thus, the basic result illustrated here
is not sensitive to resolution over the range that we have been
able to test. Cold accretion is also evident in much higher resolution SPH
simulations of the formation of individual galaxies (Katz \& Gunn 1991;
Navarro \& Steinmetz, priv. comm.) and in the Virgo Consortium's
simulations of cosmological volumes (Kay et al.\ 2000), implying
that any artifact would have to be generic to SPH,
not merely to our specific implementation of it.
However, confirmation by an independent numerical method with
better shock capturing properties is clearly desirable.
Cen \& Ostriker (1999, fig.~4) find a broad temperature distribution for
cooling radiation in their Eulerian hydro simulation, suggesting
that much of the gas in this simulation also cools without ever
reaching $T \geq 10^6\K$.
Kravtsov (priv. comm.) reports preliminary findings similar
to ours in simulations that use an adaptive refinement grid code
with a shock capturing hydrodynamics algorithm.
Present evidence, therefore, suggests that the bimodal $\tmax$
distribution is a genuine physical result, at least given the physical
assumptions of these simulations.

Our preliminary investigations suggest that these two accretion modes
are geometrically distinct, in addition to being thermally distinct.
Particles accreted in the hot mode come from a quasi-spherical
distribution, as envisioned in the conventional galaxy formation scenario.
Particles accreted in the cold mode, by contrast, travel to galaxies
along filamentary ``highways.''  Our tentative picture, therefore, is
that the first gas to enter filaments is only mildly shock heated;
as it moves along the filaments towards their nodes of intersection,
it is heated by adiabatic compression or by further
mild shocks, but the relatively slow heating and short cooling times
allow the gas to radiate its energy as quickly as it is gained.
Nagai \& Kravtsov (2002) show, in a simulation with no radiative cooling,
that the core of a filament can indeed have much lower entropy than the
outer regions.
Gas must dissipate a large amount of gravitational potential energy
before it can join an object as dense as a galaxy, but filamentary
accretion allows it to do so without ever reaching a high temperature.

The existence of an efficient, low temperature, filamentary accretion mode
could have important implications for the origin of galaxy angular momenta and
for the rapid decline of the cosmic star formation rate (SFR) at $z<1$
(Madau et al.\ 1996).
If a galaxy accretes much of its baryonic mass along filamentary structures,
then the specific angular momentum distribution of its gas and stars
may have little direct connection to that of its parent dark matter halo,
even if the total specific angular momenta are of the same order.
In the traditional picture of gas accretion, the decline of the cosmic SFR
is attributed mainly to longer cooling times in the hotter, lower
density halos that form at lower redshift.
However, it is not clear that this effect is strong enough in itself
to produce the observed sharp drop in the cosmic SFR
from $z=1$ to $z=0$ (e.g., Baugh et al.\ 1998;
Somerville \& Primack 1999; the observations themselves have significant
uncertainties).
We have shown in Murali et al. (2002) that the cosmic star formation rate
closely tracks the smooth gas accretion rate and not the rate by which
galaxies gain gas through merging.
If filamentary accretion dominates over quasi-spherical accretion,
then the decline of the cosmic SFR may be driven largely by geometrical
effects.  At high redshift, the sizes of filaments are well matched to
those of galaxies, and they can act as efficient umbilical cords,
channeling gas to the embryonic systems at their intersections.
At low redshift, however, the cross sections of typical filaments
grow to hundreds of kpc, so they tend to deliver their gas to groups and
clusters (where it is heated in accretion shocks) rather than directing it
to individual galaxies.

Our simulations incorporate supernova feedback, but its impact is
usually mild because the energy is deposited in a dense
medium with a short cooling time.  The stellar masses of the simulated
galaxies appear to be systematically too high relative to estimates
from the observed luminosity function, a problem that is fairly generic
to hydrodynamic simulations with similar physical assumptions
(e.g., Katz et al.\ 1992; Pearce et al.\ 1999; Nagamine et al.\ 2001).
Until the origin of this discrepancy is better understood,
it is difficult to assess the importance of cold mode gas accretion
in the real universe, even if it is clearly important in the
simulations themselves.
Shaun Cole (priv. comm.) has pointed out that a ``cold mode'' of
gas accretion would also appear in a semi-analytic model of galaxy formation
if cooling were allowed to proceed unchecked in halos with low virial
temperatures, since larger galaxies could then build up by mergers of these
small systems, with much of their gas never being heated to high temperatures.
However, most semi-analytic calculations suppress this ``cold mode'' by
supernova feedback, which is assumed to be more effective in low mass halos
(Dekel \& Silk 1986).
The current numerical results suggest that cold accretion
is smooth and filamentary, in which case efficient star formation
and feedback is unlikely to suppress it.
Indeed, it is possible that feedback from supernovae or AGN activity
is actually more effective in suppressing ``hot mode'' accretion, where
the incoming gas typically has larger geometrical cross section, lower
density, and higher entropy.
If this is the case, then the relative importance of the cold mode could be
even greater than it appears in these simulations.

Clearly there is more to be understood about the physical mechanisms of
cold mode gas accretion in numerical simulations, and about its robustness
to changes in numerical resolution and hydrodynamics algorithm.
However, there is now a substantial amount of evidence, from our
own simulations and from others, that cold, filamentary accretion
makes an important contribution to the buildup of galaxies.
Further investigations of this process could lead to significant
revisions in our understanding of galaxy formation and evolution.



\end{document}